\documentclass[a4paper,french]{rnti}

\usepackage[linesnumbered,ruled,vlined, french, onelanguage]{algorithm2e}
\usepackage{color}
\usepackage{adjustbox}

\usepackage[T1]{fontenc}
\usepackage[utf8]{inputenc}
\usepackage{url}

\definecolor{gray}{rgb}{0.5, 0.5, 0.5}
\definecolor{darkgreen}{rgb}{0, 0.3, 0}

\titrecourt{Enjeux \'{e}thiques de l'IA en sant\'{e}}

\nomcourt{Fabrice Muhlenbach}

\titre{Enjeux \'{e}thiques de l'IA en sant\'{e}~: une humanisation du parcours de soin
par l'intelligence artificielle~?}

\auteur{Fabrice Muhlenbach}

\affiliation{Universit\'{e} de Lyon, UJM-Saint-Etienne, CNRS \\ Laboratoire Hubert Curien, UMR 5516\\
18 rue du Professeur Beno\^{\i}t Lauras, F-42023 Saint Etienne, FRANCE \\
E-mail: \texttt{fabrice.muhlenbach@univ-st-etienne.fr} \\
ORCID: \texttt{0000-0002-1825-4290}}

\resume{\footnote{Pr\'{e}-publication d'un article \`{a} para\^{\i}tre dans la revue
\textit{Soins Cadres}.} Envisager le recours \`{a} l'intelligence artificielle pour
une plus grande personnalisation de la prise en charge du patient et une
meilleure gestion des ressources humaines et mat\'{e}rielles peut sembler une
opportunit\'{e} \`{a} ne pas manquer. Afin de proposer une meilleure humanisation du
parcours de soin, l'intelligence artificielle est un outil que les d\'{e}cideurs du
milieu hospitalier doivent s'approprier en veillant aux nouveaux enjeux
\'{e}thiques et conflits de valeurs que cette technologie engendre.}

\begin{document}

%

\section{Introduction}

Quiconque s'est d\'{e}j\`{a} retrouv\'{e} dans un service des urgences en pleine nuit, un
nourrisson pleurant dans les bras, r\^{e}verait d'\^{e}tre tout de suite pris en charge
au lieu d'\^{e}tre confin\'{e} dans une salle avec d'autres patients dans la m\^{e}me
situation de d\'{e}tresse~: ici, \`{a} la terrible impuissance d'\^{e}tre le t\'{e}moin de la
souffrance de son enfant s'ajoute l'angoisse li\'{e}e \`{a} l'ignorance du temps \`{a}
patienter avant de se retrouver face \`{a} un m\'{e}decin. \`{A} d\'{e}faut de personnel
m\'{e}dical disponible, on aurait souhait\'{e} qu'un robot d'accueil\footnote{De tels
robots d'accueil existent, \`{a} la mani\`{e}re du robot humano\"{\i}de Pepper, adopt\'{e} par
la RATP et la SNCF, qui est charg\'{e} de renseigner les voyageurs.}  puisse nous
orienter, acc\'{e}der \`{a} notre dossier m\'{e}dical et nous aider \`{a} remplir le
questionnaire d'entr\'{e}e afin de collecter les donn\'{e}es permettant \`{a} une
intelligence artificielle de poser un pr\'{e}diagnostic. En comparant les
diff\'{e}rents profils des personnes entrant au service des urgences, ce programme
d'IA pourrait \^{e}tre en mesure de prioriser les patients, d'aider \`{a} la
logistique, d'envoyer des alertes au personnel et de fournir \`{a} chaque patient
des informations permettant de le rassurer sur son \'{e}tat et d'indiquer une
estimation r\'{e}aliste du temps d'attente avant d'\^{e}tre pris en charge.

Loin d'\^{e}tre de la science-fiction, il s'agit peut-\^{e}tre d'une r\'{e}ponse \`{a} la
situation de crise que conna\^{\i}t l'h\^{o}pital o\`{u}, faute de moyens et de ressources
humaines disponibles, la douleur et le sentiment d'abandon qu'\'{e}prouvent
certains patients les am\`{e}nent \`{a} un m\'{e}contentement qui peut parfois d\'{e}g\'{e}n\'{e}rer en
incivilit\'{e}s ou agressions envers un personnel soignant d\'{e}j\`{a} bien malmen\'{e} par
des conditions de travail \'{e}puisantes. Envisager le recours \`{a} l'intelligence
artificielle pour optimiser les flux de patients et de mat\'{e}riel, r\'{e}duire les
co\^{u}ts de mobilisations des ressources humaines et de mat\'{e}riels tout en
am\'{e}liorant l'exp\'{e}rience du patient (moins de temps d'attente, un meilleur
parcours de soin), voil\`{a} de quoi r\'{e}jouir les directeurs des soins et cadres de
sant\'{e}, mais un tel changement ne peut s'effectuer sans cons\'{e}quences \'{e}thiques.

\section{Au c\oe ur de l'\'{e}thique~: la gestion des conflits de valeurs}

Qu'est-ce que l'\'{e}thique~? D'un point de vue philosophique, l'\'{e}thique concerne
les r\'{e}flexions sur les comportements \`{a} adopter pour rendre le monde humainement
habitable, avec comme finalit\'{e} la recherche d'un id\'{e}al de soci\'{e}t\'{e} et de
conduite de l'existence. D'un point de vue pratique, le c\oe ur de l'\'{e}thique
concerne des conflits entre des valeurs. Ces valeurs --~consid\'{e}r\'{e}es comme
l'id\'{e}ologie d'un groupe d'individus~-- peuvent par exemple \^{e}tre celles
propos\'{e}es par \cite{Deonna_Tieffenbach__Petit_Traite_des_Valeurs__2018} dans le
Tableau~\ref{tab:valeurs}.

\begin{table}
\begin{center}
\begin{tabular}{|ccccc|} \hline
  l'amiti\'{e} & l'amour & l'art & l'authenticit\'{e} & la beaut\'{e} \\
  le bien-\^{e}tre & la comp\'{e}tence & la confiance & la connaissance & la cr\'{e}ativit\'{e} \\
  la dignit\'{e} & l'\'{e}galit\'{e} &  l'enfance & l'\'{e}rotisme & l'honneur \\
  l'humilit\'{e} & l'humour & l'impartialit\'{e} & l'int\'{e}ressant & la justice \\
  la libert\'{e} & le luxe & le plaisir & le pouvoir & la propri\'{e}t\'{e} \\
  le sacr\'{e} & la sant\'{e} & le savoureux & la solidarit\'{e} & le sublime \\
  la tradition & l'utilit\'{e} &  la vertu &  la vie & la vie priv\'{e}e  \\ \hline
\end{tabular}
\caption{Liste des 35 valeurs du \textit{Petit Trait\'{e} des valeurs}
\citep{Deonna_Tieffenbach__Petit_Traite_des_Valeurs__2018}} \label{tab:valeurs}
\end{center}
\end{table}

Le milieu hospitalier est d\'{e}di\'{e} dans son ensemble au service d'une m\^{e}me cause~:
le \textit{patient}. Les valeurs privil\'{e}gi\'{e}es dans le contexte m\'{e}dical sont
ainsi la \textit{vie}, la \textit{sant\'{e}} et le \textit{bien-\^{e}tre} du patient.
D'un point de vue manag\'{e}rial, les directeurs d'h\^{o}pitaux et cadres de soin ont
pour fonction d'optimiser les ressources dont ils disposent et ont la charge,
qu'elles soient humaines (personnel soignant ou fonctions de support),
mat\'{e}rielles (nombre de lits, mat\'{e}riel m\'{e}dical, m\'{e}dicaments\ldots) en cherchant
\`{a} garantir les trois valeurs pr\'{e}c\'{e}demment cit\'{e}es tout en veillant \`{a} une valeur
propre \`{a} leur fonction~: l'\textit{utilit\'{e}}, c'est-\`{a}-dire la gestion et
l'organisation de ces ressources avec efficacit\'{e}.

Avec la r\'{e}volution num\'{e}rique, c'est-\`{a}-dire le fait de coder toute forme
d'information par des nombres
\citep{Serres__Petite_Poucette__2012,Berry__La_numerisation_du_monde__2010}, la
soci\'{e}t\'{e} s'est retrouv\'{e}e boulevers\'{e}e par le d\'{e}veloppement de technologies issues
de l'informatique et par l'arriv\'{e}e d'Internet. Parmi ces technologies,
l'intelligence artificielle (ou <<~IA~>>) est sortie des laboratoires et des
domaines d'application tr\`{e}s sp\'{e}cialis\'{e}s pour s'adapter \`{a} tout type de terrain.
Ainsi, le rapport Villani \citep{Villani__2018}, dans son focus sur <<~la sant\'{e}
\`{a} l'heure de l'IA~>>, d\'{e}bute par ces propos enthousiastes~:
<<~\textit{L'intelligence artificielle en sant\'{e} ouvre des perspectives tr\`{e}s
prometteuses pour am\'{e}liorer la qualit\'{e} des soins au b\'{e}n\'{e}fice du patient et
r\'{e}duire leur co\^{u}t --~\`{a} travers une prise en charge plus personnalis\'{e}e et
pr\'{e}dictive~-- mais \'{e}galement leur s\'{e}curit\'{e}~-- gr\^{a}ce \`{a} un appui renforc\'{e} \`{a} la
d\'{e}cision m\'{e}dicale et une meilleure tra\c{c}abilit\'{e}. Elle peut \'{e}galement contribuer
\`{a} am\'{e}liorer l'acc\`{e}s aux soins des citoyens, gr\^{a}ce \`{a} des dispositifs de
pr\'{e}diagnostic m\'{e}dical ou d'aide \`{a} l'orientation dans le parcours de soin.}~>>

\section{L'intelligence artificielle d\'{e}mystifi\'{e}e}

Pour prendre la mesure de ces changements en cours et \`{a} venir dans le milieu
m\'{e}dical, il convient de pr\'{e}ciser ce qu'est l'intelligence artificielle, ce
qu'elle peut ou ne peut pas faire, et, au-del\`{a} du domaine du possible, ce qu'il
est souhaitable ou non qu'elle fasse.

Par d\'{e}finition, l'intelligence artificielle est une discipline de
l'informatique qui a pour objet de simuler par ordinateur des processus de la
pens\'{e}e. L'IA se propose d'imiter sur une machine des comportements qui, si on
les rencontre, sont qualifi\'{e}s d'intelligents. Il existe ainsi de multiples
approches sur l'intelligence artificielle \citep{Russell_Norvig__IA__2010},
sch\'{e}matiquement r\'{e}parties suivant quatre domaines en fonction de l'objectif
poursuivi~:

\begin{enumerate}
  \item des syst\`{e}mes qui pensent comme des humains~: la mod\'{e}lisation cognitive qui
cherche \`{a} \'{e}tudier la nature de la pens\'{e}e humaine~;
  \item des syst\`{e}mes qui agissent comme des humains~: les programmes qui sont capables
de passer le test de Turing, c'est-\`{a}-dire de faire croire qu'il s'agit d'\^{e}tres
humains et non de machines~;
  \item des syst\`{e}mes qui pensent rationnellement~: h\'{e}riti\`{e}re des travaux de Blaise
Pascal ou Charles Babbage sur la pens\'{e}e logique, cette approche voit la pens\'{e}e
comme une forme de calcul~;
\item des syst\`{e}mes qui agissent rationnellement.
\end{enumerate}

Alors que la premi\`{e}re approche, celle de la mod\'{e}lisation cognitive, est
d\'{e}sign\'{e}e par l'expression <<~intelligence artificielle forte~>>, la derni\`{e}re,
celle des agents artificiels rationnels, n'a ni la pr\'{e}tention ni l'objectif de
s'int\'{e}resser \`{a} la pens\'{e}e ou de prendre l'\^{e}tre humain comme mod\`{e}le. C'est
pourtant cette derni\`{e}re approche, appel\'{e}e <<~intelligence artificielle
faible~>>, qui a vraiment fait conna\^{\i}tre l'IA au grand public \`{a} partir des
ann\'{e}es 2010, avec le superordinateur \textit{Watson} d'IBM qui avait r\'{e}ussi \`{a}
battre les champions humains du jeu \textit{Jeopardy!} en 2011, ou le programme
\textit{AlphaGo} de \textit{Google DeepMind} qui avait largement remport\'{e} de
nombreuses parties du jeu de go face aux joueurs humains champions d'Europe en
2015 ou du monde en 2016.

Pour comprendre comment un programme est capable de r\'{e}aliser une t\^{a}che aussi
complexe que de jouer \`{a} un jeu intellectuel (les \'{e}checs, le jeu de go) mieux
qu'un \^{e}tre humain ou d'identifier, \`{a} partir d'une image d'une tache sur la peau
ou d'un grain de beaut\'{e}, s'il s'agit de tumeurs malignes ou de taches b\'{e}nignes,
et ceci avec de meilleurs taux de succ\`{e}s que les professionnels de la sant\'{e}, il
est n\'{e}cessaire de donner quelques notions de base sur la mani\`{e}re dont sont
\'{e}labor\'{e}s de tels programmes. Un programme n'est rien d'autre qu'un algorithme
mis sous une forme compr\'{e}hensible par une machine. Un algorithme est tout
simplement le d\'{e}coupage d'une certaine op\'{e}ration en t\^{a}ches \'{e}l\'{e}mentaires, comme
une recette de cuisine qui permet d'\'{e}laborer un plat ou un dessert (une mousse
au chocolat) \`{a} partir d'ingr\'{e}dients (des \oe ufs, du sucre, du beurre et du
chocolat) pr\'{e}sents en une certaine quantit\'{e} suivant un certain ordre bien
d\'{e}fini (casser les \oe ufs, s\'{e}parer les blancs des jaunes, monter les blancs en
neige, faire fondre le chocolat avec le beurre, etc.) ou comme un proc\'{e}d\'{e}
math\'{e}matique, telle que la r\'{e}solution d'une \'{e}quation du second degr\'{e} qui
n\'{e}cessite le calcul du discriminant pour trouver les \'{e}ventuelles racines
solutions de l'\'{e}quation.

Un programme ne fait rien d'autre que suivre les instructions cod\'{e}es par le
programmeur~: on lui fournit des valeurs en entr\'{e}e (par exemple les valeurs
$a$, $b$ et $c$ de l'\'{e}quation $a x^2 + b x + c = 0$) et le programme effectue
des actions en retour (le calcul des racines solutions). La particularit\'{e} d'un
programme d'intelligence artificielle dot\'{e} de ce que l'on appelle
<<~l'apprentissage automatique~>> (ou \textit{machine learning}) est de
parvenir \`{a} changer certains param\`{e}tres en fonction des exemples qui lui sont
pr\'{e}sent\'{e}s afin d'am\'{e}liorer ses r\'{e}ponses. En pr\'{e}sentant \`{a} certains programmes
d'apprentissage automatique des millions et des millions d'images de chats et
d'images sur lesquelles il n'y a pas de chat, le syst\`{e}me parviendra de lui-m\^{e}me
\`{a} extraire des caract\'{e}ristiques pertinentes des images et \`{a} modifier ses
param\`{e}tres afin de r\'{e}pondre avec justesse <<~chat~>> \`{a} des images apprises,
mais aussi \`{a} effectuer une g\'{e}n\'{e}ralisation lui permettant de r\'{e}pondre
correctement s'il y a ou non la pr\'{e}sence de chat sur des images non apprises.

\`{A} l'heure actuelle, les programmes les plus efficaces sur ce genre de probl\`{e}me
se basent sur des mod\`{e}les appel\'{e}s <<~r\'{e}seaux de neurones artificiels~>> (parce
que leur fonctionnement copie certaines caract\'{e}ristiques des neurones
biologiques organis\'{e}s en r\'{e}seaux) et sur un proc\'{e}d\'{e} d\'{e}nomm\'{e} <<~l'apprentissage
profond~>> (ou \textit{deep learning}), c'est-\`{a}-dire des syst\`{e}mes
d'apprentissage automatique qui reposent sur une architecture en de multiples
couches entre l'entr\'{e}e (par exemple les diff\'{e}rents pixels d'une image) et la
sortie (la d\'{e}cision <<~chat~>> ou <<~non chat~>>) avec des quantit\'{e}s
consid\'{e}rables de connexions entre ces couches, ces param\`{e}tres de connexions se
modifiant petit \`{a} petit lors de l'apprentissage d'une t\^{a}che donn\'{e}e
(c'est-\`{a}-dire forcer le syst\`{e}me \`{a} r\'{e}pondre <<~chat~>> quand l'image d'un chat
est bien pr\'{e}sente, et pas autrement). Au-del\`{a} de la reconnaissance d'objets sur
des images, ces syst\`{e}mes d'apprentissage automatique vont s'appliquer \`{a} tout
type de probl\`{e}me, pourvu qu'il y ait suffisamment de donn\'{e}es \'{e}tiquet\'{e}es pour
faire un apprentissage --~c'est-\`{a}-dire que des exemples doivent \^{e}tre associ\'{e}s \`{a}
une \'{e}tiquette \`{a} apprendre fournies par un expert~-- en <<~dig\'{e}rant~>> des
informations issues d'un tr\`{e}s grand nombre de variables pour \'{e}tablir des
corr\'{e}lations entre ces variables, en faisant l'hypoth\`{e}se d'un lien causal entre
ces variables en cas de corr\'{e}lation.

Ce proc\'{e}d\'{e} d'apprentissage --~qui n'a rien de magique, il s'agit juste de
math\'{e}matique~-- ne fonctionne pas \`{a} tous les coups. Des variables peuvent en
effet \^{e}tre fortement corr\'{e}l\'{e}es entre elles, positivement ou n\'{e}gativement, pour
d'autres raisons~: on remarque que si on se r\'{e}veille apr\`{e}s avoir dormi avec ses
chaussures aux pieds, on a tr\`{e}s souvent mal \`{a} la t\^{e}te. Il n'y a pas de lien
causal \`{a} rechercher entre ces deux variables (est-ce que le fait de passer une
nuit sans enlever ses chaussures aurait un r\^{o}le perturbateur dans la
circulation du sang qui induirait des maux de t\^{e}te au r\'{e}veil~?), ces deux
\'{e}v\'{e}nements \'{e}tant simplement tous deux la cons\'{e}quence d'un autre ph\'{e}nom\`{e}ne~:
s'endormir apr\`{e}s avoir trop bu d'alcool (et donc ne plus \^{e}tre en mesure
d'enlever ses chaussures en se couchant et avoir la gueule de bois au r\'{e}veil).
N\'{e}anmoins, les d\'{e}couvertes de liens possibles entre certaines variables prises
parmi une multitude par des proc\'{e}d\'{e}s automatiques a permis de cr\'{e}er des
syst\`{e}mes de reconnaissance ou des mod\`{e}les pr\'{e}dictifs dont les performances
n'avaient jusqu'alors jamais \'{e}t\'{e} atteintes~: un changement quantitatif a permis
de r\'{e}aliser un v\'{e}ritable saut qualitatif.

Historiquement, ce changement n'a pu se faire qu'\`{a} travers des avanc\'{e}es dans
trois domaines au cours des d\'{e}cennies 2000 et 2010~:

\begin{enumerate}
  \item la disponibilit\'{e} en donn\'{e}es servant d'exemples lors de la phase
d'apprentissage,
  \item le d\'{e}veloppement de nouveaux algorithmes d'apprentissage capables de
traiter de telles quantit\'{e}s de donn\'{e}es,
  \item  la mise au point de mat\'{e}riel \'{e}lectronique (tels les processeurs graphiques)
permettant d'effectuer les calculs n\'{e}cessaires \`{a} cet apprentissage en un temps
raisonnable.

\end{enumerate}

Les programmes d'intelligence artificielle, bien que parvenant \`{a} r\'{e}soudre
certaines t\^{a}ches complexes avec brio, ne font pas vraiment preuve
d'intelligence et ne comprennent absolument pas ce qu'ils font
\citep{Dessalles__Des_intelligences_TRES_artificielles__2019}. Malgr\'{e} cela, il
est possible de donner \`{a} une machine une notion du <<~sens~>> associ\'{e} aux mots
gr\^{a}ce \`{a} une technique appel\'{e}e <<~le plongement lexical~>> (en anglais,
\textit{word embedding}). Cette technique permet de repr\'{e}senter chaque mot d'un
dictionnaire par un vecteur de nombres r\'{e}els \`{a} travers l'analyse statistique
d'un grand nombre de textes afin de retrouver les termes qui apparaissent dans
des contextes similaires et, \`{a} partir de l\`{a}, d'en d\'{e}duire des significations
apparent\'{e}es. Gr\^{a}ce \`{a} cette repr\'{e}sentation vectorielle des mots, sur laquelle
peuvent s'appliquer des op\'{e}rations arithm\'{e}tiques, il est possible de proc\'{e}der \`{a}
des raisonnements par analogie~: par exemple, la repr\'{e}sentation vectorielle du
mot <<~roi~>> moins celle du mot <<~homme~>> plus celle du mot <<~femme~>>
donne la repr\'{e}sentation vectorielle du mot <<~reine~>>.

\section{Les entreprises de l'intelligence artificielle}

Les g\'{e}ants du num\'{e}rique (en particulier les GAFAM\footnote{\textit{Google},
\textit{Apple}, \textit{Facebook}, \textit{Amazon} et \textit{Microsoft}.}  et
NATU\footnote{\textit{Netflix}, \textit{Airbnb}, \textit{Tesla} et
\textit{Uber}.} am\'{e}ricains ou les BATX chinois\footnote{\textit{Baidu},
\textit{Alibaba}, \textit{Tencent} et \textit{Xiaomi}.}), disposant de
quantit\'{e}s ph\'{e}nom\'{e}nales de donn\'{e}es sur leurs utilisateurs (le \textit{big
data}), sont aussi les entreprises leaders dans le domaine de l'intelligence
artificielle parce que l'efficacit\'{e} d'un programme d'apprentissage d\'{e}pend de la
quantit\'{e} et de la qualit\'{e} des exemples qu'on lui fournit. Il n'est par
cons\'{e}quent gu\`{e}re surprenant de voir une entreprise comme \textit{Google
LLC}/\textit{Alphabet Inc.}, par le biais de sa filiale \textit{Calico},
chercher \`{a} occuper le terrain de la sant\'{e} de fa\c{c}on disruptive en misant sur la
convergence des nanotechnologies, des biotechnologies, de l'informatique et des
sciences cognitives avec comme objectif de <<~tuer la mort~>>. Les donn\'{e}es sont
devenues un enjeu crucial dans le domaine de l'intelligence artificielle, et
les croisements d'informations issues de sources multiples am\`{e}nent \`{a} des
questions majeures dans le domaine \'{e}thique au sujet de la pr\'{e}servation de la
vie priv\'{e}e. \`{A} titre d'exemple, la soci\'{e}t\'{e} de biotechnologie \textit{23andMe},
qui propose une analyse du code g\'{e}n\'{e}tique aux particuliers, a pour cofondatrice
celle qui fut l'\'{e}pouse du cofondateur de \textit{Google}\ldots Quand on sait
qu'\textit{Alphabet Inc.} (la maison m\`{e}re de \textit{Google}), investissant
massivement dans la recherche en intelligence artificielle (avec
\textit{DeepMind Technologies}), dispose d\'{e}j\`{a} de tant de sources d'informations
(historiques du moteur de recherche, e-mails de la messagerie \textit{Gmail},
syst\`{e}me d'exploitation mobile \textit{Android} avec ses fonctions de
g\'{e}olocalisation, outils de bureautique en ligne \textit{Google Docs},
\textit{Sheets} ou \textit{Slides}, etc.), il y a de quoi vraiment s'inqui\'{e}ter
si cette entreprise poss\`{e}de en plus d'un acc\`{e}s \`{a} nos informations g\'{e}n\'{e}tiques~!

\section{IA, donn\'{e}es et cons\'{e}quences \'{e}thiques}

En tant que cadre du milieu hospitalier, quelle attitude adopter si une
entreprise priv\'{e}e nous propose un syst\`{e}me permettant d'assurer une meilleure
sant\'{e} de nos patients en \'{e}change des donn\'{e}es sur ces derniers~? Il y a l\`{a}
affrontement entre les valeurs de \textit{sant\'{e}} et de \textit{vie priv\'{e}e}.

L'intelligence artificielle et la robotisation am\`{e}nent au remplacement de
t\^{a}ches humaines r\'{e}p\'{e}titives par des automates mat\'{e}riels ou logiciels. Les
ressources humaines du syst\`{e}me de sant\'{e} seront fortement impact\'{e}es par ce
changement, et en particulier les fonctions supports (administration, gestion,
finances, fonctions logistiques g\'{e}n\'{e}rales ou m\'{e}dicotechniques)
\citep{Gruson_Kirchner__Ethique_numerique_IA__2019}. Sur le sujet des attentes
et des impacts du d\'{e}veloppement de l'intelligence artificielle dans les
h\^{o}pitaux, le cabinet de conseil EY et le CHRU de Nancy ont publi\'{e} r\'{e}cemment le
r\'{e}sultat d'une enqu\^{e}te qui fait ressortir que l'IA est tr\`{e}s majoritairement
consid\'{e}r\'{e}e comme un enjeu par les directeurs et pr\'{e}sidents de commissions
m\'{e}dicales d'\'{e}tablissements de CHU, que l'IA suscite de fortes attentes des
professionnels mais \'{e}galement des craintes et des interrogations, en
particulier au sujet du risque de d\'{e}shumanisation du travail et de la perte des
liens sociaux \citep{EY__2019}.

Un autre risque est li\'{e} aux caract\'{e}ristiques propres de ces syst\`{e}mes
artificiels. Les programmes d'intelligence artificielle am\'{e}liorent leurs
performances \`{a} partir d'un apprentissage statistique. Ils d\'{e}tectent les
co\"{\i}ncidences dans les donn\'{e}es apprises et les extrapolent sur des donn\'{e}es non
apprises. Cela a pour cons\'{e}quence de renforcer les st\'{e}r\'{e}otypes rencontr\'{e}s
implicitement dans les usages en cours. \`{A} la diff\'{e}rence de l'image d'un chat
qui pr\'{e}sente une forme prototypique et inamovible de petit f\'{e}lin \`{a} la t\^{e}te
l\'{e}g\`{e}rement arrondie et aux oreilles triangulaires, la soci\'{e}t\'{e} \'{e}volue, les
pratiques m\'{e}dicales changent. Les syst\`{e}mes artificiels apprennent du pass\'{e} pour
pr\'{e}dire le futur, en tenant compte d'un tr\`{e}s grand nombre de variables, sans
indiquer les raisons explicites qui les ont amen\'{e}s \`{a} conclure de telle ou telle
mani\`{e}re. Le r\'{e}sultat d'un apprentissage effectu\'{e} sur des donn\'{e}es d\'{e}s\'{e}quilibr\'{e}es
(c'est-\`{a}-dire quand il y a une surrepr\'{e}sentation d'une situation par rapport \`{a}
une autre) am\`{e}nera un renforcement des situations les plus pr\'{e}sentes, et ceci
de mani\`{e}re caricaturale. Un programme d'IA d'aide au tri des candidatures
employ\'{e} par l'entreprise \textit{Amazon}, qui avait effectu\'{e} son apprentissage
sur des CV re\c{c}us pendant une dizaine d'ann\'{e}es, a d\^{u} \^{e}tre abandonn\'{e} car il
discriminait les candidates, qu'il orientait sur des postes de secr\'{e}tariat,
alors que les candidatures d'hommes \'{e}taient dirig\'{e}s vers des postes de cadres.
Aux \'{E}tats-Unis, dans le domaine juridique, des logiciels d'aide \`{a} la d\'{e}cision
cens\'{e}s mieux informer les juges et homog\'{e}n\'{e}iser la mani\`{e}re dont la justice est
rendue ont aussi montr\'{e} des biais raciaux, en indiquant des risques de r\'{e}cidive
plus \'{e}lev\'{e}e ou en proposant des peines plus lourdes pour les citoyens
afro-am\'{e}ricains que pour ceux \`{a} la peau blanche. Des biais racistes ou sexistes
\'{e}mergent aussi des textes a priori neutres (des pages \textit{Wikipedia} ou des
articles de \textit{Google Actualit\'{e}s})~: en utilisant la technique du
plongement lexical et en faisant le calcul sur les repr\'{e}sentations vectorielles
des mots <<~m\'{e}decin~>> $-$ <<~homme~>> $+$ <<~femme~>>, on obtient la valeur du
mot <<~infirmi\`{e}re~>> et non celle de <<~femme m\'{e}decin~>>\ldots

\section{Conclusion}

\`{A} l'heure o\`{u} le syst\`{e}me de sant\'{e} fran\c{c}ais traverse une crise, l'introduction de
techniques issues de l'intelligence artificielle doit \^{e}tre vue comme une
opportunit\'{e}. La r\'{e}volution num\'{e}rique am\`{e}ne son lot de mutations
professionnelles et soci\'{e}tales, tout comme le firent la m\'{e}canisation et
l'automation industrielle. L'arriv\'{e}e des robots et de l'intelligence
artificielle doit \^{e}tre vue positivement, leur utilisation doit se faire au
service de l'humain, et les choix de valeurs \`{a} d\'{e}fendre sont de v\'{e}ritables
enjeux civilisationnels. Alors que la vision des \'{E}tats-Unis est de consid\'{e}rer
que les donn\'{e}es sont la propri\'{e}t\'{e} des entreprises qui les collectent \`{a} des fins
commerciales, que celle de la Chine est d'employer les donn\'{e}es de ses citoyens
\`{a} des fins de contr\^{o}le \`{a} travers le syst\`{e}me de cr\'{e}dit social, la France et
l'Union europ\'{e}enne voient dans les donn\'{e}es de la vie priv\'{e}e une ressource \`{a}
prot\'{e}ger, comme le montre l'adoption et l'application du R\`{e}glement g\'{e}n\'{e}ral sur
la protection des donn\'{e}es (RGPD).

Les directeurs des soins et cadres de sant\'{e} ne doivent pas d\'{e}roger \`{a} leurs
fonctions manag\'{e}riales d\'{e}di\'{e}es aux valeurs essentielles de \textit{vie}, de
\textit{sant\'{e}} et de \textit{bien-\^{e}tre} des patients, en recherchant
l'\textit{utilit\'{e}} dans la mani\`{e}re dont ils g\`{e}rent les ressources dont ils ont
la charge, mais ils doivent aussi \^{e}tre les garants de la protection d'autres
valeurs qui risquent d'\^{e}tre menac\'{e}es par la r\'{e}volution num\'{e}rique~: la
pr\'{e}servation de la \textit{vie priv\'{e}e} et des donn\'{e}es sensibles des patients
couvertes par le secret m\'{e}dical, la \textit{comp\'{e}tence} dans l'usage de ces
technologies num\'{e}riques, la \textit{confiance} dans les syst\`{e}mes d'aide \`{a} la
d\'{e}cision apport\'{e}s par l'IA, la \textit{connaissance} par la compr\'{e}hension des
raisons qui ont amen\'{e} le programme d'IA \`{a} proposer tel ou tel r\'{e}sultat et que
celui-ci pourra ou non suivre en toute conscience (transparence des
algorithmes, impartialit\'{e} des d\'{e}cisions, non-discrimination), et le
\textit{pouvoir} en bannissant les approches prescriptives et en laissant la
ma\^{\i}trise \`{a} l'utilisateur humain afin de demeurer un acteur \'{e}clair\'{e} et ma\^{\i}tre de
ses choix (que ce soit le consentement du patient ou la validation par un
expert humain d'un pr\'{e}diagnostic pos\'{e} par un programme d'IA).

Pour un directeur des soins ou cadre de sant\'{e}, ce n'est qu'\`{a} travers la
combinaison de la bonne connaissance de ce qu'est l'outil IA, de ses
possibilit\'{e}s et de ses limites, et de son engagement dans le respect des
valeurs fondamentales du patient, que l'intelligence artificielle pourra \^{e}tre
employ\'{e}e \`{a} des fins d'une meilleure humanisation du parcours de soin.

\Fr

\bibliographystyle{rnti}
\bibliography{Muhlenbach__Soins_Cadres__2020}

\providecommand\Fr{}
\providecommand\Eng{}
\providecommand\andname{and}
\providecommand\andnamec{and}

\begin{thebibliography}{}


\bibitem[{Berry}(2010){Berry}]{Berry__La_numerisation_du_monde__2010}
Berry, G. (2010).
\newblock {\em La numérisation du monde}.
\newblock Paris:  Éditions De Vive Voix.

\bibitem[{Deonna \andnamec{} Tieffenbach}(2018){Deonna \andnamec{}
  Tieffenbach}]{Deonna_Tieffenbach__Petit_Traite_des_Valeurs__2018}
Deonna, J. \andname{} E.~Tieffenbach (Eds.) (2018).
\newblock {\em Petit Traité des valeurs}.
\newblock Collection <<~Science \& métaphysique~>>. Paris:  Éditions d'Ithaque.

\bibitem[{Dessalles}(2019){Dessalles}]{Dessalles__Des_intelligences_TRES_artificielles__2019}
Dessalles, J.-L. (2019).
\newblock {\em Des intelligences TRÈS artificielles}.
\newblock Paris:  Éditions Odile Jacob.

\bibitem[{{EY} \andnamec{} {CHRU Nancy}}(2019){{EY} \andnamec{} {CHRU
  Nancy}}]{EY__2019}
{EY} \andname{} {CHRU Nancy} (2019).
\newblock Baromètre de maturité de l'{IA} dans les {CHU}.
\newblock \\  {\scriptsize \url{http://www.chru-nancy.fr/images/colloqueIA/1911SG412_Barometre_AI_CHU_Nancy-VF3.pdf}}.

\bibitem[{Gruson \andnamec{} Kirchner}(2019){Gruson \andnamec{}
  Kirchner}]{Gruson_Kirchner__Ethique_numerique_IA__2019}
Gruson, D. \andname{} C.~Kirchner (2019).
\newblock Éthique, numérique et intelligence artificielle~: quels enjeux pour
  les cadres de santé~?
\newblock {\em Soins Cadres\/}~{\em 28\/}(115), 47--49.
\newblock \\  {\scriptsize \url{https://doi.org/10.1016/j.scad.2019.09.021}}.

\bibitem[{Russell \andnamec{} Norvig}(2010){Russell \andnamec{}
  Norvig}]{Russell_Norvig__IA__2010}
Russell, S.~J. \andname{} P.~Norvig (2010).
\newblock {\em Intelligence artificielle\/} (3 ed.).
\newblock Montreuil: Pearson France.

\bibitem[{Serres}(2012){Serres}]{Serres__Petite_Poucette__2012}
Serres, M. (2012).
\newblock {\em Petite Poucette}.
\newblock Collection <<~Manifestes~>>. Paris: Éditions Le Pommier.

\bibitem[{Villani}(2018){Villani}]{Villani__2018}
Villani, C. (2018).
\newblock Donner un sens à l'intelligence artificielle~: pour une stratégie
  nationale et européenne.
\newblock {Mission confiée par le Premier Ministre Édouard Philippe}, Mission
  parlementaire du 8 septembre 2017 au 8 mars 2018.
\newblock  \\ {\scriptsize \url{https://www.aiforhumanity.fr/pdfs/9782111457089_Rapport_Villani_accessible.pdf}}.

\end{thebibliography}

\end{document}